\documentclass[preprint,12pt]{elsarticle}
\usepackage{epsfig}%
\usepackage{amsmath}%
\usepackage{graphicx}

\begin{document}

\begin{frontmatter}
\title{Negative Binomial Distribution \\ and the multiplicity moments at the LHC}
\author{Michal Praszalowicz\fnref{fn1}}
\address{M. Smoluchowski Institute of Physics, Jagellonian University, \\
Reymonta 4, 30-059 Krakow, Poland} \fntext[fn1]{e-mail: {\tt
michal@if.uj.edu.pl}}

\begin{abstract}
In this work we show that the latest LHC data on multiplicity moments
$C_2-C_5$ are well described by a two-step model in the form of a
convolution of the Poisson distribution with energy-dependent source
function. For the source function we take $\Gamma$ Negative Binomial
Distribution. No unexpected behavior of Negative Binomial Distribution
parameter $k$ is found. We give also predictions for the higher energies
of 10 and 14 TeV.
\end{abstract}
\end{frontmatter}
\vspace{2cm}

 One of the widely discussed, yet unsolved, problems in high
energy hadron scattering is the production mechanism of low and medium
$p_{\text{T}}$ hadrons. Here perturbative QCD cannot be applied and one
resorts to phenonemonological models and/or Monte\ Carlo generators
\cite{Kittel}. With the advent of any new hadron accelerator the
quantities first studied are charged particle multiplicities. Indeed, the
first physics LHC paper, the one by Alice collaboration \cite{:2009dt},
dealt with the average multiplicity. By now data on multiplicity moments
were published by Alice \cite{Aamodt:2010ft} and CMS
\cite{Khachatryan:2010nk}. It is therefore important to find if their
behavior can be explained in terms of some simple phenomenological models
or the widely used MC generators. The latter has been recently addressed
in Refs.\cite{Fialkowski:2010mh}.

Throughout this paper we explore the well known observation, recently
recalled in Ref.\cite{Bialas:2010zza}, that multiparticle production can
be described by the probability distribution $P(n)$ which is a
superposition of some unknown distribution of sources $F$, and the Poisson
distribution describing particle emission from one source. This is a
typical situation in many microscopic models of multiparticle production.
For example in Dual Parton Model (DPM) particles are emitted by chains
spanned between the colliding protons (for review see
Ref.\cite{Capella:1992yb}). In that case there are many sources whose
number increases with energy. Similarly in Quark-Gluon String Model (QGSM)
emission proceeds from $n$ cut-pomerons (Ref.\cite{Kaidalov:2009hn} and
references therein), and again is assumed to be Poissonian.

Here we refrain from formulating a microscopic multiparticle production
model and assume a simple phenomenological formula that captures, however,
the above mentioned physics of independent emissions encoded in DPM or
QGSM:
\begin{equation}
P(n)=%
{\displaystyle\int\limits_{0}^{\infty}}
dt\,F(t)\,e^{-\bar{n}t}\frac{(\bar{n}t)^{n}}{n!}. \label{convol}%
\end{equation}
Here $t$ is a fraction of the average multiplicity, and $F(t)$ the
distribution of sources that contribute fraction $t$ to the multiplicity
probability $P(n)$. Normalization conditions require%
\begin{equation}%
{\displaystyle\int\limits_{0}^{\infty}}
dt\,F(t)=%
{\displaystyle\int\limits_{0}^{\infty}}
dt\,t\,F(t)=1.\label{normaliz}%
\end{equation}

There are two useful properties of Eq.(\ref{convol}) that will be of
importance throughout this paper. The first one is the fact that
\emph{factorial} moments of multiplicity distribution measure directly the
$F_{m+1} $ moments of the
source:%
\begin{equation}
\left\langle n(n-1)(n-2)\ldots(n-m)\right\rangle =\bar{n}^{m+1}%
{\displaystyle\int\limits_{0}^{\infty}}
dt\,t^{m+1}F(t)=\bar{n}^{m+1}F_{m+1}.\label{factmoms}%
\end{equation}
Because of (\ref{normaliz}) average multiplicity%
\begin{equation}
\left\langle n\right\rangle =\bar{n}.
\end{equation}
Factorial moments can be expressed through scaled regular moments%
\begin{equation}
C_{m}=\frac{\left\langle n^{m}\right\rangle }{\left\langle n\right\rangle
^{m}}%
\end{equation}
that have been measured at the LHC \cite{Aamodt:2010ft,Khachatryan:2010nk}.
For the first five moments we have:%
\begin{align}
C_{2}  & =\frac{1}{\left\langle n\right\rangle }+F_{2},\nonumber\\
C_{3}  & =3\frac{C_{2}}{\left\langle n\right\rangle }-2\frac{1}{\left\langle
n\right\rangle ^{2}}+F_{3},\nonumber\\
C_{4}  & =6\frac{C_{3}}{\left\langle n\right\rangle }-11\frac{C_{2}%
}{\left\langle n\right\rangle ^{2}}+6\frac{1}{\left\langle n\right\rangle
^{3}}+F_{4},\nonumber\\
C_{5}  & =10\frac{C_{4}}{\left\langle n\right\rangle }-35\frac{C_{3}%
}{\left\langle n\right\rangle ^{2}}+50\frac{C_{2}}{\left\langle n\right\rangle
^{3}}-24\frac{1}{\left\langle n\right\rangle ^{4}}+F_{5}.\label{Cn}%
\end{align}

The second property of Eq.(\ref{convol}) is that for large multiplicities
(\emph{i.e.} for large energies) it implies an approximate KNO (Koba,
Nielsen and Olesen) scaling \cite{Koba:1972ng}. Indeed, in the limit
$\bar{n}\rightarrow\infty$ and fixed $n/\bar{n}$ one can apply the saddle
point approximation to calculate $dt$ integral in
(\ref{convol}) leading to~\cite{Bialas:2010zza}%
\begin{equation}
\psi\equiv\bar{n}P(n)\simeq F\left(  \frac{n}{\bar{n}}\right)  .\label{KNOpsi}%
\end{equation}
KNO scaling says that function $\psi$ depends only on $\tau=n/\bar{n}$.

KNO scaling is seen approximately in the multiplicity distributions
measured at SPS and higher energies (see for review
\cite{GrosseOetringhaus:2009kz}) including the LHC
\cite{Aamodt:2010ft,Khachatryan:2010nk}. This fact gives strong
justification for formula (\ref{convol}) which also allows to give
definite predictions for the violation of the KNO scaling. Originally KNO
scaling has been derived assuming Feynman scaling \cite{Feynman:1969ej},
which states that the central rapidity density saturates at asymptotic
energies. The latter is clearly not seen in the data ({\em e.g.}
\cite{:2009dt}), on the contrary central rapidity density grows as a power
of energy. For example for $\left\vert \eta\right\vert <0.5$
\cite{McLerran:2010ex}:
\begin{equation}
\left.  \frac{dn}{d\eta}\right\vert _{\left\vert \eta\right\vert <0.5}%
\sim0.755\left(  \frac{W}{1\,\text{GeV}}\right)  ^{0.23}\label{mult}%
\end{equation}
where $W=\sqrt{s}$. For constant $\bar{n}$ all multiplicity moments would be
constant as well.

Here we see the advantage of the convolution model (\ref{convol}) since it
implies approximate KNO scaling also for energy dependent $\bar{n}$. This
energy dependence introduces in turn energy dependence of the moments, as
clearly seen from equations (\ref{Cn}), even if function $F$ is energy
independent. Unfortunately this dependence alone would contradict the data
since for constant $F_{n}$ multiplicity moments decrease with energy (for
growing $\left\langle n\right\rangle $).

Therefore the source function $F$ has to depend on energy and its moments
have to win over the decrease generated by the multiplicity growth through
Eqs.(\ref{Cn}). In Ref.\cite{Bialas:2010zza} the method of recovering $F$
from the data has been discussed, without, however, reference to the
recent measurements at the LHC. In DPM or QGSM violation of the KNO
scaling proceeds by an increase of the number of sources (chains,
pomerons) with increasing energy.

Here, rather than constructing a microscopic model of multiparticle
production, we choose the explicit form of $F(t)$ and check whether we are
able to describe multiplicity moments measured by Alice and CMS. To this
end we choose for $F$ negative binomial distribution (NBD)
\cite{GrosseOetringhaus:2009kz}:%
\begin{equation}
F(t,k)=\frac{k^{k}}{\Gamma(k)}t^{k-1}e^{-kt}\label{NBD}%
\end{equation}
which is known to describe relatively well the data at lower energies
\cite{Kittel}. Distribution (\ref{NBD}) depends on one parameter $k$,
which -- as explained above -- has to depend on $W$. It is known from the
analysis of lower energy data that, depending on energy, $k\sim 4$~-~2 and
decreases with increasing energy. Let us remind that for $k=1$ the
probability distribution $P_{\text{NBD}}$ is given by geometrical
distribution $P(n)=\left\langle n\right\rangle ^{n}/(1+\left\langle
n\right\rangle )^{n+1}$. With increasing $k$ ($1/k\rightarrow0$), the
distribution $P_{\text{NBD}}$ is getting narrower tending to the Poisson
distribution.

Based on experimental evidence of the wide occurrence of NBD, several
possible explanations have been proposed in the literature (for review see
Ref.\cite{Kittel}). The NBD has been mostly interpreted in terms of
(partial) stimulated emissions or cascade processes
\cite{Giovannini:1985mz}. More recently NBD has been derived from the
Color Glass Condensate (CGC) approach giving explicit prediction for the
energy dependence of parameter $k$ at high energies being of the order of
the LHC energy range \cite{Gelis:2009wh}. Here, contrary to the lower
energy trend, parameter $k$ is expected to grow with energy, as it is
directly connected to the saturation scale which increases with energy.
Similar behavior is found in String Percolation Model (SPM)
\cite{deDeus:2010id}, where -- once percolation is achieved -- $k$ starts
to grow with energy like in the CGC. It is therefore interesting to see if
the new regime of growing $k$ has been already achieved at the LHC, which
is one of the motivations behind the present work.

For negative binomials%
\begin{equation}
F_{m+1}=\frac{k(k+1)\ldots(k+m)}{k^{m+1}}.
\end{equation}
The first equation of (\ref{Cn}) gives:%
\begin{equation}
C_{2}=\frac{1}{\left\langle n\right\rangle }+1+\frac{1}{k}\quad\rightarrow
\quad\frac{1}{k}=C_{2}-1-\frac{1}{\left\langle n\right\rangle }.\label{1overk}%
\end{equation}
Using (\ref{1overk}) we get for higher moments%
\begin{align}
C_{3}  & =C_{2}(2C_{2}-1)-\frac{C_{2}-1}{\left\langle n\right\rangle
},\nonumber\\
C_{4}  & =C_{2}(6C_{2}^{2}-7C_{2}+2)-2\frac{3C_{2}^{2}-4C_{2}+1}{\left\langle
n\right\rangle }+\frac{C_{2}-1}{\left\langle n\right\rangle ^{2}},\nonumber\\
C_{5}  & =C_{2}(24C_{2}^{3}-46C_{2}^{2}+29C_{2}-6)-2\frac{18C_{2}^{3}%
-34C_{2}^{2}+19C_{2}-3}{\left\langle n\right\rangle }\nonumber\\
& +\frac{14C_{2}^{2}-23C_{2}+9}{\left\langle n\right\rangle ^{2}}-\frac
{C_{2}-1}{\left\langle n\right\rangle ^{3}}.\label{C35}%
\end{align}

\begin{figure}[h]
\centering
\includegraphics[scale=0.63]{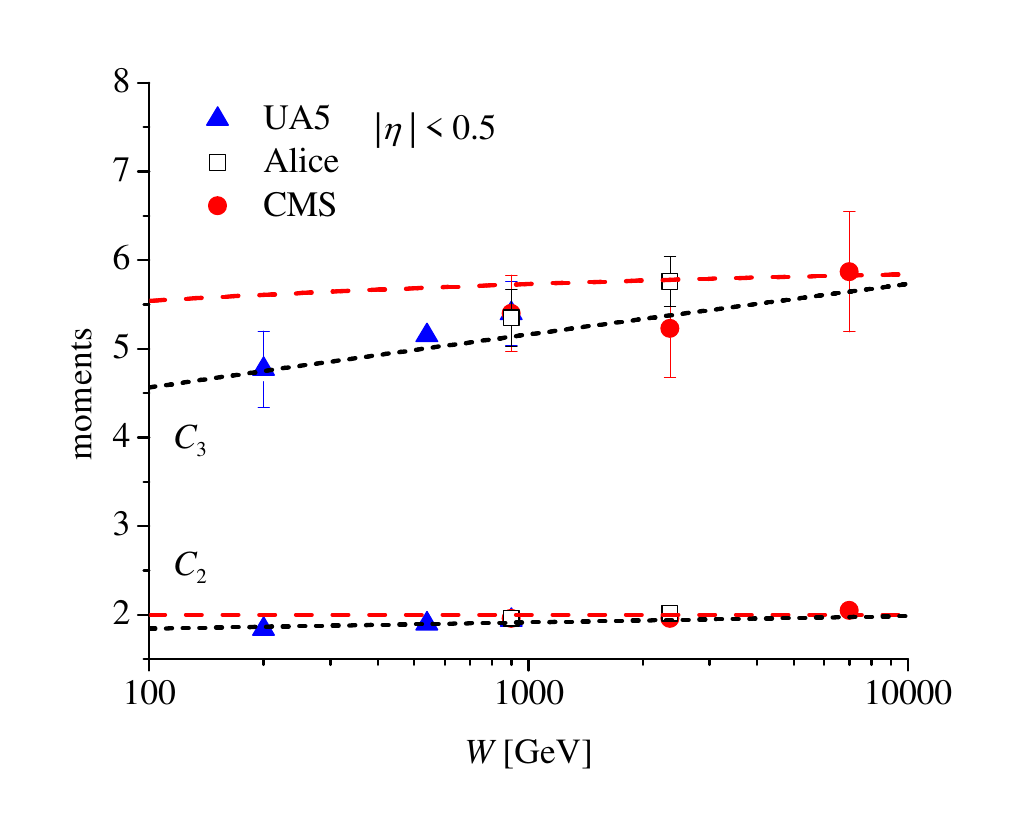}
\includegraphics[scale=0.63]{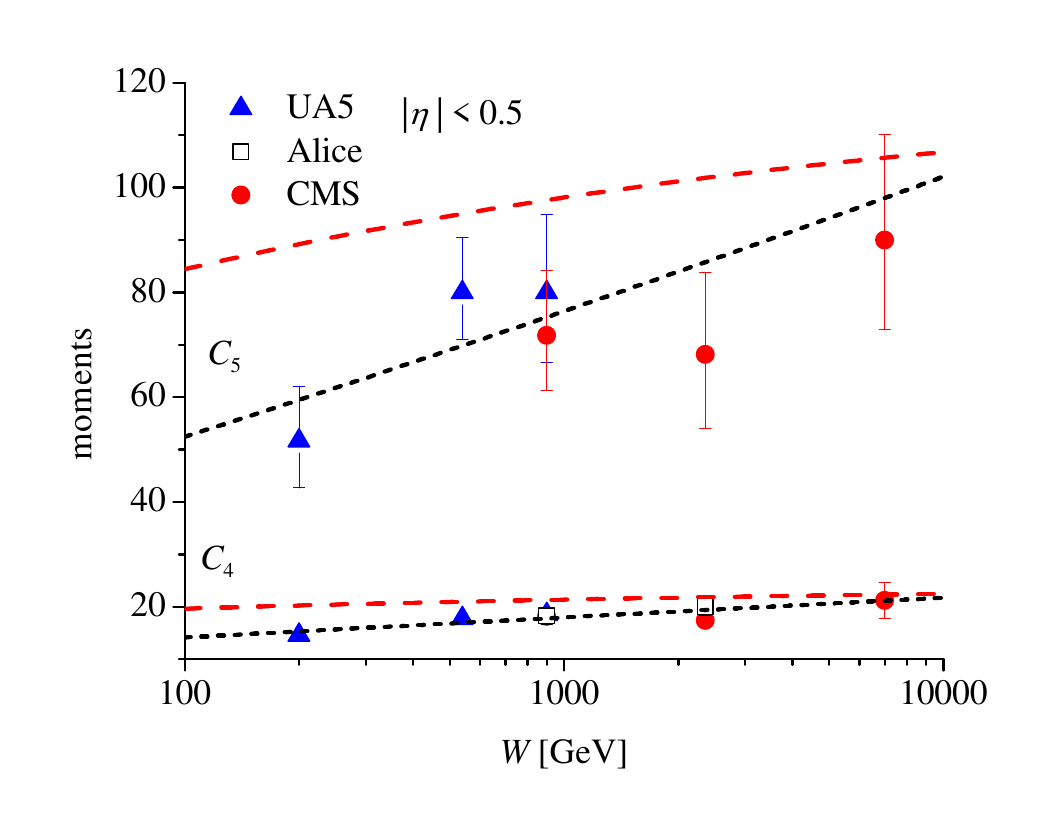}\newline%
\includegraphics[scale=0.63]{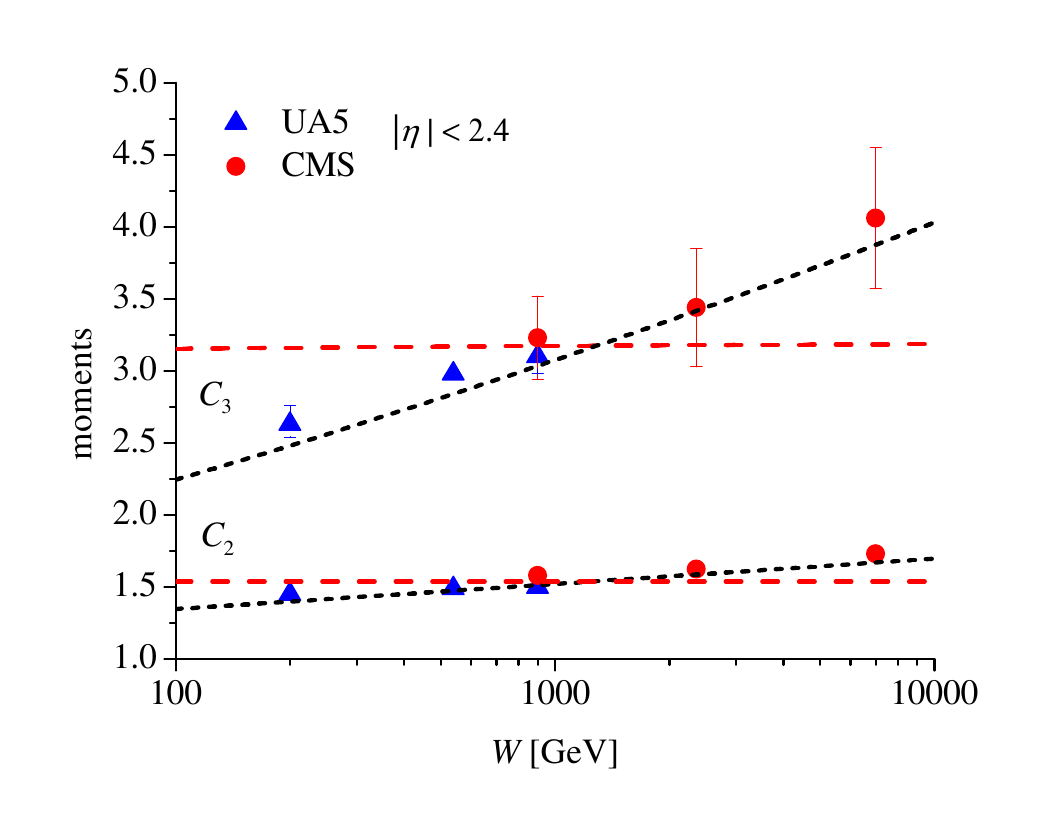}
\includegraphics[scale=0.63]{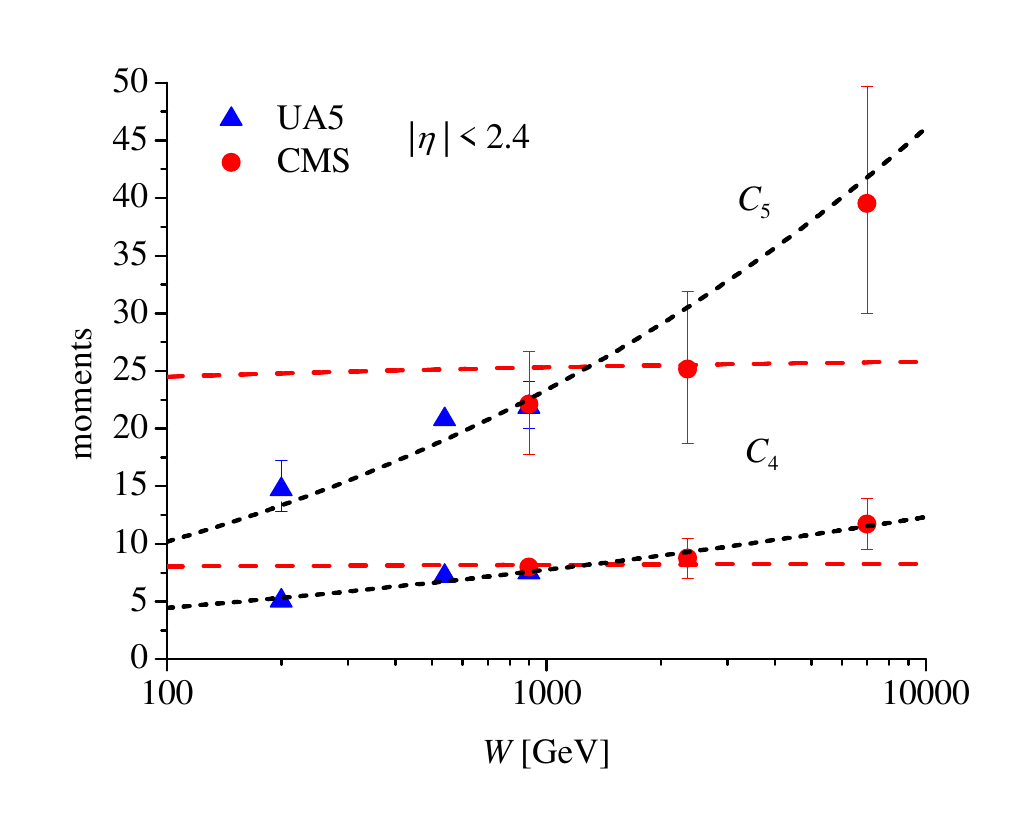}
\caption{Multiplicity moments measured
by UA5 \cite{Alner:1985zc}, Alice \cite{Aamodt:2010ft} and
CMS \cite{Khachatryan:2010nk} for two rapidity intervals in function of energy
$W=\sqrt{s}$. Long dash (red) lines correspond to constant $C_{2}$ moments
equal to 2 and 1.54 for two rapidity intervals $|\eta|<0.5$ and $|\eta|<2.4$
respectively. Short dash (black) lines correspond to energy dependent $C_{2}$
as explained in the text. }%
\label{C2C5}%
\end{figure}

Let us first observe that for constant $C_{2}$, which is approximately
true, at least for $\left\vert \eta\right\vert <0.5$ where $C_{2}\simeq2$,
higher moments grow with energy. This is depicted in Fig.\ref{C2C5} where
long dash (red) line corresponds to constant $C_{2}$ in two rapidity
intervals $\left\vert \eta\right\vert <0.5$ ($C_{2}\simeq2$) and
$\left\vert \eta\right\vert <2.4$ ($C_{2}\simeq1.54$). It is clearly seen
that for negative binomial distribution used here this growth is, however,
too slow. For larger rapidity intervals $\Delta\eta$ multiplicity is also
larger $dn/d\eta=0.755\,\Delta\eta\,W^{0.23}$ and therefore the inverse
powers of multiplicity are less important than for smaller rapidity
ranges. Moreover for $\left\vert \eta\right\vert <2.4$ second moment
$C_{2}\simeq1.54$ and therefore the coefficients in front of powers
$\left\langle n\right\rangle ^{-m}$ are also smaller than for $\left\vert
\eta\right\vert <0.5$. Therefore, as seen in Fig.\ref{C2C5}, for
$\left\vert \eta\right\vert <2.4$ in the fit where $C_2=$const., all other
moments are nearly constant as well.

\begin{table}[ptb]
\centering
\begin{tabular}
[c]{|cc|cc|}\hline
Fig. & $\left\vert \eta\right\vert <\eta_{0}$ & $a$ & $b$\\\hline
\ref{C2C5} & $0.5$ & $1.702$ & $0.071$\\
\ref{C2C5} & $2.4$ & $0.997$ & $0.175$\\\hline
\ref{CAlice} & $0.5$ & $1.597$ & $0.111$\\
\ref{CAlice} & $1.0$ & $1.377$ & $0.118$\\
\ref{CAlice} & $1.3$ & $1.219$ & $0.149$\\\hline
\end{tabular}
\caption{Values of parameters $a$ and $b$ of Eq.(\ref{C2fit}) used in
Figs.\ref{C2C5} and \ref{CAlice}.}%
\label{Tabab}%
\end{table}

\begin{figure}[h]
\centering
\includegraphics[scale=0.50]{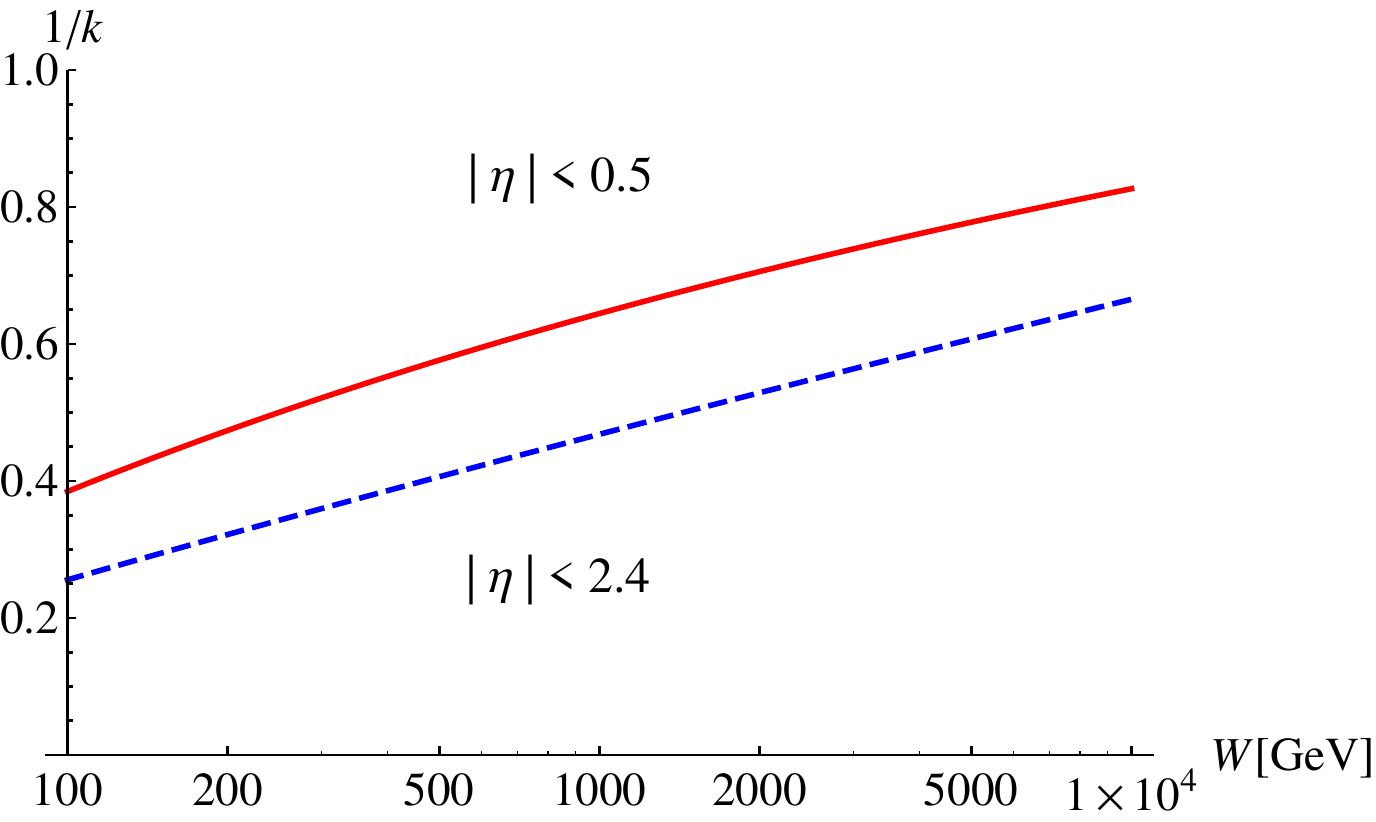}
\caption{Growth of $1/k$ used in Fig.\ref{C2C5}%
.}%
\label{kmin1}%
\end{figure}

In order to reproduce the growth seen in the data we have therefore to
require a mild increase of $C_{2}$ with $W$. Higher moments $C_{m}$ are
proportional to higher powers of $C_{2}^{m}$ and should therefore grow
faster with $W$ with increasing $m$. This trend is clearly seen in the
data. To this end we choose
to approximate $C_{2}$ by a linear function of $\log W$:%
\begin{equation}
C_{2}=a+b\log(W\text{[GeV]}).\label{C2fit}%
\end{equation}
In order to find parameters $a$ and $b$ we choose to fit $C_{4}$ rather
than $C_{2}$. In both cases $\left\vert \eta\right\vert <0.5$ and
$\left\vert \eta\right\vert <2.4$ moment $C_{4}$ grows rather fast with
$W$ having still reasonable errors. Fitting $C_{2}$ gives usually too slow
increase of higher moments, whereas fitting $C_{4}$ reproduces all moments
with good precision. This is easily seen from Fig.\ref{C2C5}. The
parameters of the fit (\ref{C2fit}) are given in Table \ref{Tabab} and the
resulting energy dependence of $1/k$ is plotted in Fig. \ref{kmin1}. We
see that the trend from lower energies continues: $k$ decreases with
energy but is still rather far from $k=1$. For $\eta<0.5$ and $W=0.9$, 7
and 14 TeV, $k=1.58$, 1.25 and 1.18 respectively. Should this dependence
continue, $k=1$ would be reached for $W\sim250$~TeV. On the other hand
dependence of $k$ on rapidity is quite pronounced.

With parametrizations (\ref{mult}) and (\ref{C2fit}) we are able to
predict the first moments for higher energies at which the LHC will be
running in the future. The results are displayed in Table \ref{Tabpred}.

\begin{table}[ptb]
\centering
\begin{tabular}
[c]{|c|cc|cc|}\hline
$W$ [TeV] & 10 & 14 & 10 & 14\\
& \multicolumn{2}{c|}{$|\eta|<0.5$} & \multicolumn{2}{c|}{$|\eta|<2.4$%
}\\\hline
$\left\langle n\right\rangle $ & 6.28 & 6.78 & 31.61 & 34.15\\
$C_{2}$ & 1.98 & 1.99 & 1.70 & 1.72\\
$C_{3}$ & 5.73 & 5.81 & 4.03 & 4.18\\
$C_{4}$ & 21.74 & 22.30 & 12.33 & 13.11\\
$C_{5}$ & 102.11 & 106.02 & 46.07 & 50.39\\\hline
\end{tabular}
\caption{Predictions for multiplicity moments.}%
\label{Tabpred}%
\end{table}

\begin{figure}[h]
\centering
\includegraphics[scale=0.63]{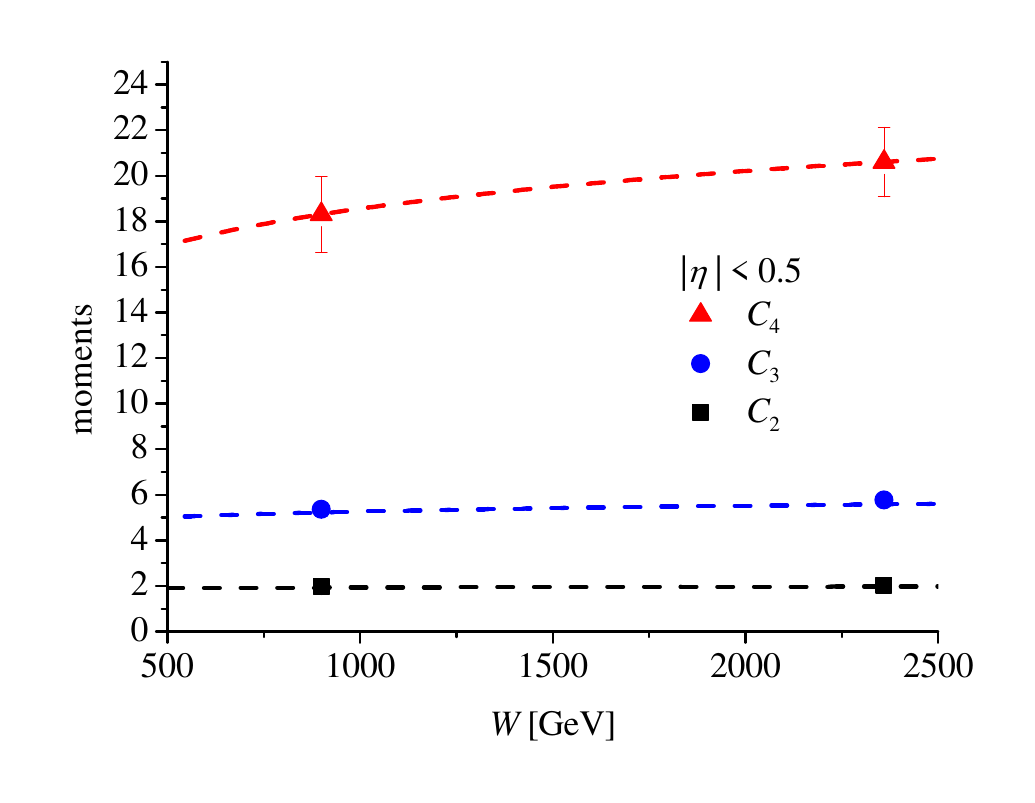}
\includegraphics[scale=0.63]{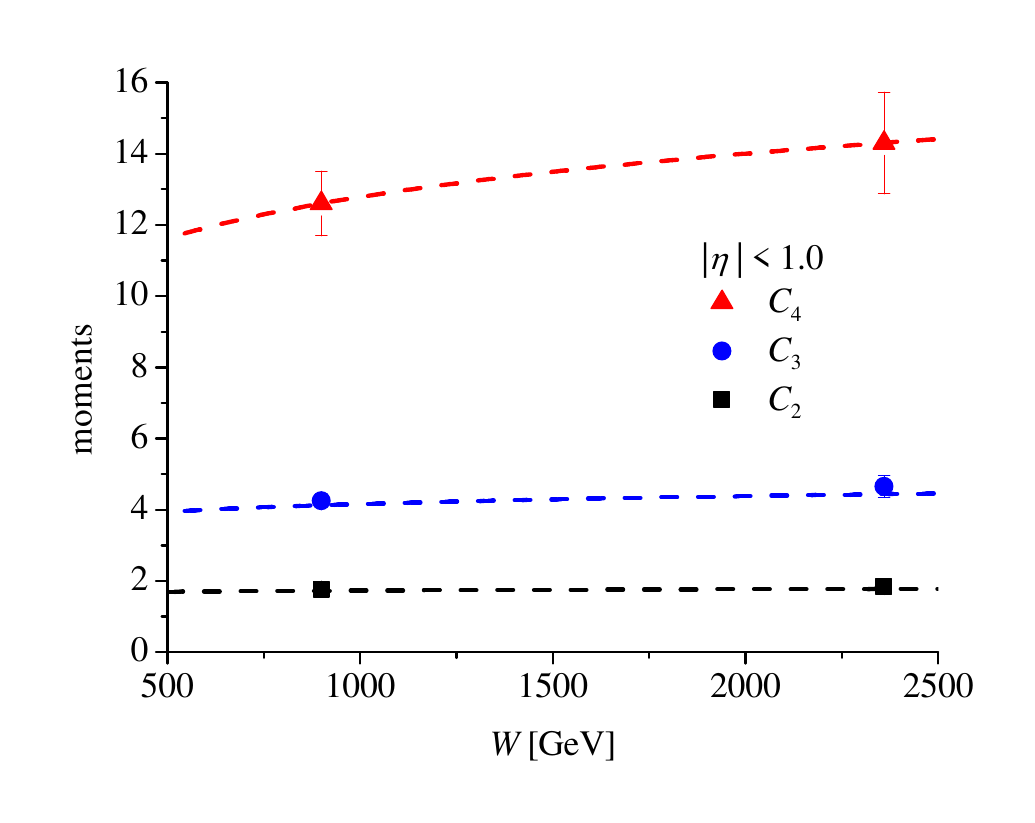}\newline%
\includegraphics[scale=0.63]{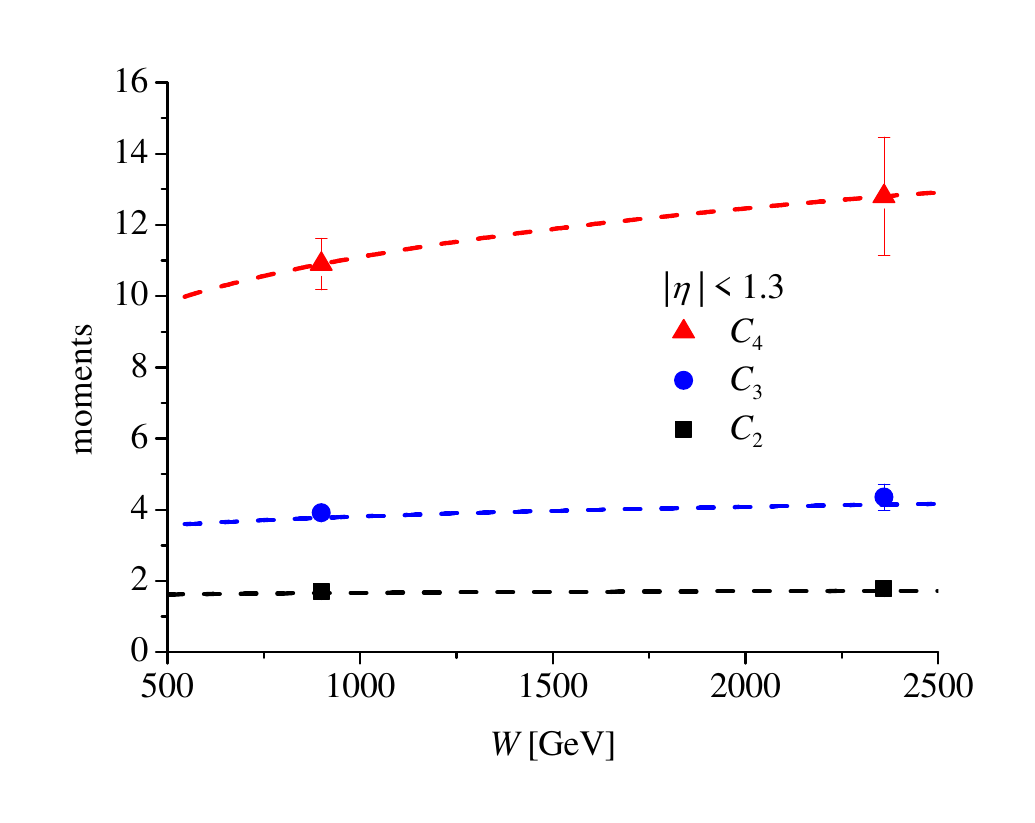}
\caption{Multiplicity moments in
different rapidity intervals as measured by Alice \cite{Aamodt:2010ft}
together with NBD fit with
$C_{2}$ parametrized as in Eq.(\ref{C2fit}) and Table \ref{Tabab}. }%
\label{CAlice}%
\end{figure}

Alice collaboration published results for the multiplicity moments
$C_{2}-C_{4}$ for two energies: 0.9 and 7 TeV and three rapidity intervals
$|\eta|<0.5$, 1 and 1.3. We repeated the same procedure described above
for the Alice data fitting $C_{4}$ with the help of Eq.(\ref{C2fit}). The
resulting parameters are collected in Table \ref{Tabab} and the moments
are plotted in Fig.\ref{CAlice}. We see good agreement of NBD fits for all
three rapidity intervals.

To conclude: we have used the convolution model (\ref{convol}) with
distribution of sources given by negative binomial function (\ref{NBD}) to
fit multiplicity moments measured recently by Alice and CMS collaborations
at the LHC. We have shown that convolution model implies that normalized
$C_{m}$ multiplicity moments decrease with increasing energy as inverse
powers of the average multiplicity (\ref{1overk},\ref{C35}) if the
distribution function $F(t)$ is energy independent. Such a behavior
contradicts data. Assuming NBD for $F(t)$ and logarithmic growth
(\ref{C2fit}) of $C_{2}$ moment, we have been able to reproduce the
multiplicity moments over the wide range of energies for different
rapidity intervals. The input growth of $C_{2}$ with energy can be easily
translated to a decrease of the parameter $k$ of NBD function (\ref{NBD}).
This behavior is consistent with lower energies and does not exhibit the
change predicted by CGC \cite{Gelis:2009wh} and/or SPM
\cite{deDeus:2010id}. We also made predictions for higher energies which
will be soon accessible at the LHC. Unfortunately we are still lacking a
microscopic model explaining energy dependence of $k$ parameter of the NBD
distribution.

\vspace{1cm}

\noindent{\bf Acknowledgments:} I am grateful to Andrzej Bialas for
discussions that triggered this research and for reading the manuscript.
Also discussions with Krzysztof Fialkowski were of considerable help.
Comments by Carlos Pajares and Elena Kokoulina are also acknowledged.

\end{document}